\documentclass[11pt]{article}
\pdfoutput=1
\usepackage[latin1]{inputenc}
\usepackage{amsmath}
\usepackage{amsfonts}
\usepackage{amssymb}
\usepackage{pstricks}
\usepackage{amsthm}
\usepackage{mathrsfs}
\usepackage{amssymb}
\usepackage{cancel}
\usepackage{slashed}
\usepackage{graphicx}
\usepackage[font=small,labelfont=bf]{caption}
\usepackage{float}
\usepackage{csquotes}
\usepackage[bottom]{footmisc} 
\usepackage{tikz}
\usepackage{setspace}
\usepackage{bbding}
\usepackage{cite}
\usetikzlibrary{matrix,arrows,decorations.pathmorphing}
\usepackage{framed}
\usepackage{color}
\usepackage{wrapfig}
\definecolor{shadecolor}{RGB}{224,238,238}
\newcommand{\nn}{\nonumber}

\makeatletter

\@addtoreset{equation}{section}
\makeatother

\def\lsim{\;\raise0.3ex\hbox{$<$\kern-0.75em\raise-1.1ex\hbox{$\sim$}}\;}
\def\gsim{\;\raise0.3ex\hbox{$>$\kern-0.75em\raise-1.1ex\hbox{$\sim$}}\;}
\def\beq{\begin{equation}}   \def\eeq{\end{equation}}
\def\ba{\begin{array}}       \def\ea{\end{array}}
\def\bea{\begin{eqnarray}}   \def\eea{\end{eqnarray}}
\def\nn{\nonumber}

\theoremstyle{definition} 
\date{\today}

\begin{document}

\begin{titlepage}
\begin{flushright}
LPT Orsay 18-91
\end{flushright}


\begin{center}

\begin{doublespace}

\vspace{1cm}
{\Large\bf Gradient flows for $\beta$ functions via multi-scale renormalization group
equations} 
\vspace{2cm}

{\bf{Ulrich Ellwanger$^{a}$}}\\
\vspace{1cm}
{\it  $^a$ Laboratoire de Physique Th\'eorique, UMR 8627, CNRS, Universit\'e de Paris-Sud, Universit\'e
Paris-Saclay, 91405 Orsay, France}

\end{doublespace}

\end{center}
\vspace*{2cm}
\begin{abstract}

Renormalization schemes and cutoff schemes allow for the introduction of
various distinct renormalization scales for distinct couplings. 
We consider the coupled renormalization group
flow of several marginal couplings which depend on just as many renormalization
scales. The usual $\beta$ functions describing the flow with respect to a common
global scale are assumed to be given. Within this framework one can always
construct a metric and a potential in the space of couplings such that the
$\beta$ functions can be expressed as gradients of the potential. Moreover
the potential itself can be derived explicitely from a prepotential which,
in turn, determines the metric. Some examples of renormalization group flows are considered, and the metric and the potential are compared to expressions obtained elsewhere.

\end{abstract}

\end{titlepage}

\newpage
\section{Introduction}
\label{sec:intro}

Originally multi-scale renormalization group (RG) flows were introduced to deal with
physical problems involving distinct energy scales \cite{Einhorn:1983fc}.
On the other hand it is plausible to consider multi-scale RG flows motivated by
purely formal arguments: 

In dimensional regularization marginal couplings
(i.e. dimensionless in $d=4$) acquire a dimension $d-4$ which requires the introduction
of a scale $\mu$, and in perturbation theory the corresponding renormalized
couplings depend on $t\equiv \log(\mu^2/\mu_0^2)$ where $\mu_0$ serves to
define initial conditions for the running couplings.
In the presence of several marginal couplings $g_a$, $a=1\dots n_g$, it is standard
to introduce a single scale $\mu$ common to all couplings, since this allows to
construct RG equations
for Green functions with respect to an overall change of scale.
However, {\it a priori} it is allowed and possible to
introduce as many parameters $\mu_i$ or $\tau_i\equiv \log(\mu_i^2/\mu_{0i}^2)$,
$i=1...n_g$. An overall change of scale can still be defined provided
all $\tau_i$ are related to an overall scale $t$.

In the presence of an ultraviolet (UV) cutoff $\Lambda$ the renormalization group can
also be used to describe the running of bare couplings with $\Lambda$ keeping the
renormalized couplings fixed. A UV cutoff $\Lambda$ must not necessarily be universal:
Consider, for example, a momemtum space cutoff of propagators which decrease rapidly for
$p^2 > \Lambda^2$. A priori it is possible to chose different cutoffs for
different fields. Although the number of fields (counting multiplets as single fields)
does not necessarily coincide with the number of marginal couplings 
one obtains again the possibility to introduce $n_g$ parameters $\tau_i$
now defined as $\tau_i\equiv \log(\Lambda_i^2/\mu_0^2)$. Distinct momentum space cutoffs
can also be introduced in the form of distinct form factors attached to the vertices
corresponding to marginal couplings, as it happens automatically in the
case of compositeness. Actually the so-called gradient flow in field space
(not to be confused with the here considered gradient flow for couplings/$\beta$
functions), originally introduced for gauge fields on a lattice \cite{Luscher:2009eq},
serves also as a UV cutoff for correlation functions of composite operators and
could be generalized to distinct cutoffs for distinct couplings.
Finally Pauli-Villars regularization allows for several distinct cutoffs as well.

Subsequently we will use
the idea of $n_g$ scales $\tau_i$ independently from whether these refer to
renormalization points $\mu_i$ or to UV cutoffs $\Lambda_i$.

Computing the radiative corrections to vertices associated to $n_g$ marginal
couplings the various couplings and scales will mix at least in higher loop order.
Consequently, in general each coupling $g_a$ will depend on each scale $\tau_i$
leading to a system of $\beta$ functions
\beq\label{eq:1.1}
\beta_a^i(g) \equiv \frac{\partial g_a}{\partial \tau_i}\; .
\eeq
Assuming as many couplings $g_a$ as scales $\tau_i$ and linearly independent $\beta_a^i(g)$
this set of partial derivatives
can formally be inverted to give $\frac{\partial \tau_i}{\partial g_a}(g)$.

On the other hand it remains possible to define a universal overall scale (or a cutoff)
$t$ with respect to which the properties of a physical system change unless it is
scale invariant. Varying $t$ the
couplings $g_a$ satisfy standard (although scheme dependent) RG equations
$\frac{\partial g_a}{\partial t}=\beta_a(g)$.
We will assume that the scales $\tau_i$ are proportional to $t$
such that
\beq\label{eq:1.2}
\frac{d\tau_i}{dt}\equiv \frac{\partial \tau_i}{\partial g_a}\frac{\partial g_a}{\partial t}
\equiv \frac{\partial \tau_i}{\partial g_a}\beta_a(g) 
= C_i
\eeq
where the constants $C_i$ may differ from $1$ for different scales $\tau_i$.
But since these drop out (cancel) in the interesting quantities below we will
consider $C_i=1$.

It is the aim of the present paper to show that the concept of different
scales $\tau_i$ leads naturally to the definition of a gradient flow
\beq\label{eq:1.3}
\eta^{a b}(g)\beta_b(g) = \frac{\partial \Phi(g)}{\partial g_a}\; .
\eeq
In addition we find that the potential $\Phi(g)$ is related to a prepotential $P$ via
\beq\label{eq:1.4}
\Phi(g)=\frac{dP(g(t))}{dt}= \beta_a \frac{\partial P(g)}{\partial g_a}\; .
\eeq
In principle such a prepotential can always be constructed if one solves the system
of coupled RG equations for $g_a(t)$, inserts the solutions into
the potential $\Phi(g(t))$, integrates
with respect to $t$ and re-expresses $t$ in terms of $g_a(t)$. In practice these steps
are hardly feasable, whereas within the present approach the prepotential is
related to the metric $\eta^{ab}$ (see the next section) which allows for its construction.

The possibility to express $\beta$ functions in terms of a metric $\eta^{a b}(g)$
and a potential $\Phi(g)$ was observed first by Wallace and Zia
\cite{Wallace:1974dx,Wallace:1974dy} for a multi-component $\varphi^4$ theory.
The consideration of Weyl consistency conditions
for local couplings in a gravitational background in dimensional regularization
led Osborn and Jack to explicit expressions for a metric $\eta^{a b}(g)$
and a potential $\Phi(g)$ \cite{Osborn:1989td,Osborn:1991gm,Jack:1990eb,Jack:2013sha,Jack:2015tka};
the symmetry of the metric matrix is possibly spoiled, however,
in higher order in perturbation theory.

A candidate $\eta_Z^{a b}$ for a metric is the correlation function of two composite
operators\break
$l^{2d}\left< O^a(x) O^b(0) \right>|_{|x|=l}$ ($l$ denotes an UV cutoff) 
where the composite operators $O^a$, $O^b$ are dual to the couplings $g_a$, $g_b$
respectively. Such a metric was introduced by Zamolodchikov \cite{Zamolodchikov:1986gt}
in order to show the irreversibility of the RG flux in
$d=2$ dimensional field theory where the positivity of $\eta_Z^{a b}$ can be shown.

It turned out to be difficult to demonstrate the irreversibility of the
RG flow in $d=4$ \cite{Cardy:1988cwa,Mavromatos:1988zq,Osborn:1989td,Jack:1990eb,
Shore:1990wq, Cappelli:1990yc,Cappelli:1991ke,Dolan:1993vh,Osborn:1999az,
Komargodski:2011vj,Luty:2012ww,Fortin:2012hn,Antipin:2013pya}. In particular
there remains the possibility of limit cycles \cite{Morozov:2003ik,Fortin:2012cq},
i.e. recurrent trajectories related to non-vanishing $\beta$ functions.
Such field theories are nevertheless conformal but the irreversible flow
concerns functions which differ from $\beta$ functions \cite{Fortin:2012hn}.

Couplings $g_a$ can be considered as sources for composite operators $O^a$, at
least if promoted to local quantities $g_a(x)$. Then a functional $G(g_a)$ can
be defined such that derivatives of $G(g_a)$ with respect to $g_a$ generate
correlation functions of operators $O^a$ \cite{Zamolodchikov:1990wx}. This
allows to relate the Zamolodchikov metric $\eta_Z^{a b}\sim \left< O^a O^b \right>$
to the second derivative of $G$, $\eta_Z^{a b}\sim \frac{\partial^2 G}{\partial g_a
\partial g_b}$. We are not very precise here since, within the present framework of
multiple scales, we find a somewhat different expression for the metric
$\eta^{a b}$ in \eqref{eq:1.3}.

The starting point of our approach is purely algebraic and could find applications
for RG flows beyond quantum field theory. We will compare, however,
our results for gradient flows in some simple field theory models
to those obtained elsewhere.

\section{Gradient flow from multiple scales}

As stated in the Introduction we consider $n_g$ marginal couplings $g_a$ depending
on $n_g$ scales $\tau_i$. We assume that the matrix of partial derivatives
$\frac{\partial g_a}{\partial \tau_i}(g)$ can be inverted such that
$\frac{\partial \tau_i}{\partial g_a}(g)$ exists, and that eq~\eqref{eq:1.2} holds.

We consider a prepotential $P(\tau(g))$ (omitting indices of $g_a$ and $\tau_i$
if these appear as arguments of functions); its total derivative with respect to
an overall scale $t$ will be identified with the potential $\Phi(\tau(g))$:
\beq\label{eq:2.1}
\Phi(\tau(g)) = \frac{dP(\tau(g))}{dt} = \frac{\partial P(\tau(g))}{\partial g_a} \beta_a= 
\frac{\partial P(\tau(g))}{\partial \tau_i} \frac{\partial \tau_i}{\partial g_a} \beta_a
\eeq
with
\beq\label{eq:2.2}
\beta_a = \frac{dg_a}{dt}
\eeq 
assumed to be known. Next we consider the derivative of \eqref{eq:2.1}
with respect to $g_a$:
\beq\label{eq:2.3}
\frac{\partial}{\partial g_a} \Phi(\tau(g)) = \left(\frac{\partial}{\partial g_a}
\frac{\partial P(\tau(g))}{\partial \tau_i}\right) \frac{\partial \tau_i}{\partial g_b}
\beta_b + \frac{\partial P(\tau(g))}{\partial \tau_i} \frac{\partial}{\partial g_a}
\left(\frac{\partial \tau_i}{\partial g_b} \beta_b\right)\; .
\eeq

Due to \eqref{eq:1.2} the second term on the right hand side of \eqref{eq:2.3}
vanishes. The first term on the right hand side of \eqref{eq:2.3} can be rewritten
as
\beq\label{eq:2.4}
\frac{\partial^2 P(\tau(g))}{\partial \tau_j \partial \tau_i} \frac{\partial \tau_j}{\partial g_a}
\frac{\partial \tau_i}{\partial g_b} \beta_b
\equiv \eta^{ab}\beta_b\; ,
\eeq
hence \eqref{eq:2.3} assumes the form of a gradient flow,
\beq\label{eq:2.5}
\frac{\partial}{\partial g_a} \Phi(\tau(g)) = \eta^{ab}\beta_b
\eeq
with
\beq\label{eq:2.6}
\eta^{ab} = \frac{\partial^2 P(\tau(g))}{\partial \tau_j \partial \tau_i} \frac{\partial \tau_j}
{\partial g_a} \frac{\partial \tau_i}{\partial g_b}\; .
\eeq
The metric \eqref{eq:2.6} is manifestly symmetric and covariant under
redefinitions $g \to g'(g)$. Note that $\eta^{ab}$ differs from
$\frac{\partial^2 P}{\partial g_a \partial g_b}$; the difference are terms of
the form $\frac{\partial P}{\partial \tau_i} \frac{\partial^2 \tau_i}
{\partial g_a \partial g_b}$. From \eqref{eq:2.6} positivity of the
metric depends now on the positivity of $\frac{\partial^2 P}
{\partial \tau_j \partial \tau_i}$ and properties of
$\frac{\partial \tau_i}{\partial g_a}$ on which we cannot make general statements.

Independently from the positivity of $\eta^{ab}$ the above arguments allow to formulate a
potential flow for a general system of $\beta$ functions. We obtain no constraints on
terms in the $\beta$ functions in the form of Weyl consistency conditions as in dimensional
regularization \cite{Osborn:1989td,Osborn:1991gm,Jack:1990eb,Fortin:2012hn,Jack:2013sha,
Jack:2015tka}. The explicit construction of the above gradient flow from a given set
$\beta$ functions with respect to an overall scale $t$ requires, however, to
consider some subtleties.

Given a set of $n_g$ $\beta$ functions $\beta_a$ the first task is to find $n_g$
independent solutions of~\eqref{eq:1.2} for $\tau_i(g)$,
\beq\label{eq:2.7}
 \frac{\partial \tau_i(g)}{\partial g_a}\beta_a(g) = C_i\; ,
\eeq
for nonzero constants $C_i$ which may all be taken as $1$ since a constant rescaling of
$\tau_i$ cancels in $\eta^{ab}$. If the system is not
degenerate there exist $n_g$ independent solutions for $\tau_i(g)$ which involve
arbitrary functions of $n_g-1$ expressions $\varphi_k(g)$; $\varphi_k(g)$ are
independent solutions of the set of corresponding homogeneous
($C_i=0$) equations \eqref{eq:2.7}.

In cases where the lowest order terms of $\beta_a$ are
of the form $\beta_a=b_a\; g_a^{\ n}+\dots$ 
(with $n$ an integer $\neq 1$, no sum over $a$) it is natural to take 
$\tau_i(g)=-\delta_i^a\frac{1}{b_a(n-1)}g_a^{1-n}+\dots$ such that 
$\tau_i(g)=t$ to lowest order, and to construct the
higher order terms subsequently. (If the $\beta$ functions are known to a given
order in perturbation theory it can be useful to supplement them
with formally higher order terms in $g$ to find analytic expressions
for $\frac{\partial \tau_i}{\partial g_a}$ satisfying \eqref{eq:2.7}.
Explicit expressions for $\tau_i(g)$ which require to integrate
$\frac{\partial \tau_i(g)}{\partial g_a}$ are actually never required.)
In other cases of $\beta_a$ one has some freedom in the construction of
$\frac{\partial \tau_i}{\partial g_a}$, but such redefinitions in the
space of $\tau_i$ drop out in the final quantities which depend on
$g_a$ only.

With $\frac{\partial \tau_i}{\partial g_a}(g)$ and its inverse
$\frac{\partial g_a}{\partial \tau_i}(g)$
 at hand one can proceed with
the construction of a metric $\eta^{a b}$. $\eta^{a b}$ has to satisfy
integrability conditions which can be derived as follows. Consider the
following derivatives of the prepotential $P(\tau(g))$:
\beq\label{eq:2.8}
\frac{\partial}{\partial g_a} \frac{\partial P(\tau(g))}{\partial \tau_i}
=\frac{\partial^2 P(\tau(g))}{\partial \tau_i \partial \tau_j}
\frac{\partial \tau_j}{\partial g_a} = \eta^{a b} \frac{\partial g_b}{\partial \tau_i}
\eeq
which imply the integrability conditions
\beq\label{eq:2.9}
\frac{\partial}{\partial g_c}\left(\eta^{a b} \frac{\partial g_b}{\partial \tau_i}\right)
= 
\frac{\partial}{\partial g_a}\left(\eta^{c b} \frac{\partial g_b}{\partial \tau_i}\right)\; .
\eeq
In order to solve \eqref{eq:2.9} it can be helpful to expand the derivatives such
that \eqref{eq:2.9} becomes
\beq\label{eq:2.10}
\frac{\partial \eta^{ab}}{\partial g_c}\frac{\partial g_b}{\partial \tau_i}
+\eta^{ab}\frac{\partial}{\partial g_c}\frac{\partial g_b}{\partial \tau_i}
=
\frac{\partial \eta^{cb}}{\partial g_a}\frac{\partial g_b}{\partial \tau_i}
+\eta^{cb}\frac{\partial}{\partial g_a} \frac{\partial g_b}{\partial \tau_i}\; .
\eeq
Contracting \eqref{eq:2.10} with $\frac{\partial \tau_i}{\partial g_d}$
leads to
\beq\label{eq:2.11}
\frac{\partial \eta^{ad}}{\partial g_c}-\frac{\partial \eta^{cd}}{\partial g_a}
= \eta^{cb}L^{ad}_{\phantom{ab}b} - \eta^{ab}L^{cd}_{\phantom{ab}b}
\eeq
with
\beq\label{eq:2.12}
L^{ad}_{\phantom{ab}b}=\frac{\partial \tau_i}{\partial g_d} \frac{\partial}{\partial g_a}
\frac{\partial g_b}{\partial \tau_i} = 
- \frac{\partial g_b}{\partial \tau_i} \frac{\partial^2 \tau_i}{\partial g_a \partial g_d}\; .
\eeq
In the last step we have used
\beq\label{eq:2.13}
0=\frac{\partial}{\partial g_a} \delta^d_b = \frac{\partial}{\partial g_a}
\left(\frac{\partial g_b}{\partial \tau_i} \frac{\partial \tau_i}{\partial g_d}\right)
= L^{ad}_{\phantom{ab}b} + \frac{\partial g_b}{\partial \tau_i} 
\frac{\partial^2 \tau_i}{\partial g_a \partial g_d}\; .
\eeq

Given $\frac{\partial \tau_i}{\partial g_a}(g)$ and its inverse
$\frac{\partial g_a}{\partial \tau_i}(g)$ it is straightforward to compute
$L^{ad}_{\phantom{ab}b}$ from the last term in \eqref{eq:2.12}.

Note that there are more integrability conditions \eqref{eq:2.11} than those
which follow from  \eqref{eq:2.5} alone and read
\beq\label{eq:2.14}
\frac{\partial}{\partial g_c} \left(\eta^{ab}\beta_b\right)=
\frac{\partial}{\partial g_a} \left(\eta^{cb}\beta_b\right)\; .
\eeq
However not all (symmetric) solutions $\eta^{ab}$ of \eqref{eq:2.14} guarantee
that $\eta^{ab}$ is covariant under redefinitions $g\to g'(g)$.
On the other hand this is guaranteed by solutions $\eta^{ab}$ of \eqref{eq:2.11};
it suffices to contract the last two terms in \eqref{eq:2.8} with
$\frac{\partial \tau_i}{\partial g_d}$. Once a metric satisfying
\eqref{eq:2.11} has been obtained
a potential $\Phi(g)$ can be found by integration of \eqref{eq:2.5}, and a
prepotential can be found by integration of \eqref{eq:2.8}.

Again the solutions of the system of partial differential differential equations
\eqref{eq:2.11} are not unique. In the considered cases we found no obstruction
for diagonal metrics $\eta^{ab} \sim \delta^{ab}f_a(g)$, but such ans\"atze do
not always lead to the simplest expressions for the diagonal elements $f_a(g)$ of
$\eta^{ab}$. These ambiguities are not related to redefinitions in the space of
couplings since redefinitions would also affect the $\beta$ functions; these
have been taken as fixed inputs, however. In the next Section we consider some
examples.

\section{Examples}

First we consider a system of 3 two-loop $\beta$ functions for gauge couplings
where fermion loops generate mixings at the two-loop level as in the Standard
Model. We maintain the notation $g_1$, $g_2$, $g_3$ of the previous sections
where $g_a$ are related to the usual gauge couplings $\alpha$ by $g_a=\frac{\alpha_a}{4\pi}$.
The $\beta$ functions are written as
\bea
\beta_1 &=& b_{10} g_1^2+b_{11} g_1^3 + b_{12}g_1^2 g_2 + b_{13}g_1^2 g_3\; ,\nn \\
\beta_2 &=& b_{20} g_2^2 + b_{21}g_2^2 g_1+b_{22} g_2^3 + b_{23}g_2^2 g_3\; ,\nn \\
\beta_3 &=& b_{30} g_3^2 + b_{31}g_3^2 g_1 + b_{32}g_3^2 g_2+b_{33} g_3^3\; . \label{eq:3.1}
\eea

In the Standard Model we have \cite{Machacek:1983tz}
\bea
b_{10}&=\frac{41}{6},\quad b_{11}=\frac{199}{18},\quad b_{12}=\frac{9}{2},\quad b_{13}&=\frac{44}{3},\nn\\
b_{20}&=-\frac{19}{6},\quad b_{21}=\frac{3}{4},\quad b_{22}=\frac{35}{4},\quad b_{23}&=12,\nn\\
b_{30}&=-7,\quad b_{31}=\frac{11}{6},\quad b_{32}=\frac{9}{2},\quad b_{33}&=-26\; .\label{eq:3.2}
\eea

It is fairly easy to find $\tau_i(g)$ which satisfy \eqref{eq:2.7} to the considered
order with $C_i=1$ and $\tau_i=t$ to lowest order:
\bea
\tau_1&=&-\frac{1}{b_{10}g_1}-\frac{1}{b_{10}}\left(\frac{b_{11}}{b_{10}}\log g_1+\frac{b_{12}}{b_{20}}\log g_2 + \frac{b_{13}}{b_{30}}\log g_3\right)\; ,\nn \\
\tau_2&=&-\frac{1}{b_{20}g_2}-\frac{1}{b_{20}}\left(\frac{b_{21}}{b_{10}}\log g_1+\frac{b_{22}}{b_{20}}\log g_2 + \frac{b_{23}}{b_{30}}\log g_3\right)\; ,\nn \\
\tau_3&=&-\frac{1}{b_{30}g_3}-\frac{1}{b_{30}}\left(\frac{b_{31}}{b_{10}}\log g_1+\frac{b_{32}}{b_{20}}\log g_2 + \frac{b_{33}}{b_{30}}\log g_3\right)\; .\label{eq:3.3}
\eea

The quantities $\frac{\partial \tau_i}{\partial g_a}(g)$ and $\frac{\partial g_a}{\partial \tau_i}(g)$
can now be obtained straightforwardly. The integrability conditions \eqref{eq:2.11}
admit solutions corresponding to an expansion of the metric $\eta^{ab}$ around the unit matrix:
\bea
\eta^{11}&=&1+\frac{b_{21} g_2^3+b_{31}g_3^3}{3 b_{10} g_1^2}\; ,\nn\\
\eta^{22}&=&1+\frac{b_{12} g_1^3+b_{32}g_3^3}{3 b_{20} g_2^2}\; ,\nn\\
\eta^{33}&=&1+\frac{b_{13} g_1^3+b_{23}g_2^3}{3 b_{30} g_3^2}\; .\label{eq:3.4}
\eea

With this metric one finds a potential $\Phi(g)$ of the form
\bea
&&\Phi(g)=\frac{1}{3}\Bigg(g_1^3\left(b_{10}+\frac{3}{4}b_{11}g_1+b_{12}g_2+b_{13}g_3\right)
                        +g_2^3\left(b_{20}+\frac{3}{4}b_{22}g_2+b_{21}g_1+b_{23}g_3\right)\nn \\
         &&\phantom{\Phi(g)=}+g_3^3\left(b_{30}+\frac{3}{4}b_{33}g_3+b_{31}g_1+b_{32}g_2\right)\Bigg)\; .
\label{eq:3.5}
\eea

By construction $\Phi(g)$ can be derived from a prepotential $P(g)$ as in \eqref{eq:2.1}, 
$\Phi(g)=\frac{\partial P(g)}{\partial g_a}\beta_a$, with
\beq\label{eq:3.6}
P(g)=\frac{1}{6}(g_1^2+g_2^2+g_3^2)
-\frac{1}{36}\left(
\frac{b_{11}g_1^3}{b_{10}}+
\frac{b_{22}g_2^3}{b_{20}}+
\frac{b_{33}g_3^3}{b_{30}}\right)\; .
\eeq

It is remarkable that the prepotential $P(g)$ does not depend on the mixing terms in the
$\beta$ functions.

The metric \eqref{eq:3.4} and the potential 
\eqref{eq:3.5} differ from the ones for the same system of $\beta$ functions
in \cite{Morozov:2003ik} where the potential consists in quartic terms in $g_a$ only (to
two-loop order). They differ also from the metric $\eta_{JO}$ obtained by Jack and Osborn from
Weyl consistency conditions \cite{Jack:1990eb}. In the space of gauge couplings their metric $\eta_{JO}$ is 
also diagonal, but of the form $\eta_{JO}^{aa}\sim \frac{N_a}{g_a^2}$ with constants $N_a$
to two-loop order. As a consequence consistency conditions among the two-loop terms of
the $\beta$ functions (in dimensional regularisation and minimal subtraction) can be derived,
see also \cite{Antipin:2013sga}. We found, however, that an expansion of $\eta^{ab}$
around $\eta_{JO}^{aa}$ cannot satisfy the integrability conditions \eqref{eq:2.11}.
(We recall that the metric $\eta_{JO}^{ab}$ is not guaranteed to be symmetric to
higher loop order.)
Here, on the other hand, we obtain the potential from a simple prepotential.

The other example is more involved already to one-loop order. It concerns a scalar with
quartic self interaction and a Yukawa coupling to a Fermion, like the Higgs-top
sector of the Standard Model with a quartic Higgs coupling $\lambda |H|^4$ and a top
quark Yuhawa coupling $h_t$. Our notation is 
\beq\label{eq:3.7}
g_1=\frac{h_t^2}{16\pi^2}\; ,\qquad g_2=\frac{\lambda}{16\pi^2}\; .
\eeq

The general one-loop $\beta$ functions are
\beq\label{eq:3.8}
\beta_1=a_1 g_1^2\; ,\qquad \beta_2=b_1 g_2^2+b_2 g_1 g_2 + b_3 g_1^2
\eeq
where in the Standard Model
\beq\label{eq:3.9}
a_1=\frac{9}{4}\; ,\quad b_1=12\; ,\quad b_2=6\; ,\quad b_3=-3\; .
\eeq

The general solution of eq.~\eqref{eq:2.7} (again with $C_i=1$) for $\tau_i(g)$ is of the form
\beq\label{eq:3.10}
\tau_i = -\frac{1}{a_1 g_1} + F_i(X)
\eeq
where $F_i(X)$ is an arbitrary function of
\beq\label{eq:3.11}
X=\frac{a_1}{w}\log\left(\frac{w-\alpha}{w+\alpha}\right)-\log g_1\quad
\text{where} \quad
w=\sqrt{(b_2-a_1)^2-4 b_1 b_3}\; ,\quad
\alpha=2 b_1 \frac{g_2}{g_1}+b_2-a_1\; .
\eeq
(The argument of the root $w$ is positive for $b_3 < 0,\ b_1 > 0$ as in the Standard
Model.)

We have studied various ans\"atze for $F_i(X)$ without observing substantial
differences in the final results (since related by redefinitions of $\tau_i$);
subsequently we consider the simplest possibility
\beq\label{eq:3.12}
\tau_1=-\frac{1}{a_1 g_1}\; ,\qquad \tau_2 = -\frac{1}{a_1 g_1} + X\; .
\eeq

Among the solutions of the integrability conditions \eqref{eq:2.11} for the
metric $\eta^{ab}$ we discuss the one which allow for expansions of
the potential $\Phi(g)$ and the prepotential $P(g)$ in powers of couplings
(without logarithms or dilogarithms). 
This metric is off-diagonal and, using $\beta_2$ from \eqref{eq:3.8},
can be written as
\bea
\eta^{11}&=& \frac{1}{3 a_1 g_1^5} \beta_2^3 -\frac{g_2}{2 g_1^4} \beta_2^2+
\left(\frac{3}{10}b_1^2 g_2^4-\frac{1}{6}(b_2^2+2b_1 b_3) g_1^2 g_2^2 +\frac{3}{2}b_3^2 g_1^4
\right)\frac{g_2}{g_1^4} +\frac{b_2 a_1 g_2^2}{3 g_1^2} -\frac{b_3^3 g_1}{3 a_1}\; ,\nn\\
\eta^{22}&=& \frac{1}{g_1^2}(\beta_2-a_1 g_1 g_2)(2b_1 g_2+(b_2-a_1)g_1)   \; ,\nn\\
\eta^{12}&=&\frac{1}{g_1^3}(\beta_2-a_1 g_1 g_2)(b_3 g_1^2-b_1g_2^2)   \; .\label{eq:3.13}
\eea

The corresponding potential $\Phi(g)$ is
\bea
\Phi(g)&=&\frac{\beta_2^3}{3g_1^2}-\frac{b_3^3 g_1^4}{12}
-\frac{a_1 g_2^2}{g_1}\left(\frac{4}{5}b_1^2 g_2^3+\frac{3}{2}b_1 b_2 g_1 g_2^2
+\frac{2}{3} g_2 g_1^2(b_2^2+2b_1 b_3)
+b_2 b_3 g_1^3 \right)\nn\\
&&+a_1^2 g_2^3\left(\frac{1}{2}b_1 g_2+\frac{1}{3}b_2g_1\right)\; .\label{eq:3.14}
\eea

It can be derived as in \eqref{eq:2.1} from the prepotential
\beq\label{eq:3.15}
P(g)=\frac{1}{9}(2b_1b_3+b_2^2-2b_2a_1+a_1^2)g_2^3 +\frac{1}{3}b_3(b_2-a_1)g_1g_2^2
+\frac{1}{3}b_3^2g_1^2g_2 -\frac{b_3^3 g_1^3}{36 a_1} +\frac{b_1(b_2-a_1)g_2^4}{6g_1}
+\frac{b_1^2 g_2^5}{15 g_1^2}\; .
\eeq
Note that the matching of the various coefficients in 
$\Phi(g)=\beta_1 \frac{\partial P(g)}{\partial g_1} + \beta_2 \frac{\partial P(g)}{\partial g_2}$
is highly nontrivial, and that the expression for $P(g)$ is actually somewhat simpler than
the one for $\Phi(g)$. But both expressions for the metric and the potential differ
considerably from the ones in \cite{Jack:1990eb} and \cite{Morozov:2003ik}.

\section{Conclusions}

Using the formalism of multi-scale RG equations we have shown how a potential
flow for a set of $n_g$ couplings and corresponding $\beta$ functions can be
constructed. Since the metric is not necessarily positive the flow is not
necessarily irreversible. This cannot be expected, however, since the formalism holds
equally for systems with limit cycles.

A particular feature of the present construction is that the potential $\Phi(g)$
derives always from a prepotential $P(g)$ as in \eqref{eq:1.4}, related to
the metric as in \eqref{eq:2.6}. Contracting
\eqref{eq:1.3} with $\beta_a$ and using \eqref{eq:1.4} one obtains
\beq\label{eq:4.1}
\beta_a\eta^{a b}\beta_b = \frac{d^2 P(g(t))}{dt^2}
\eeq
which may be helpful for the study of global features of the RG flow.

A holographic formulation of the RG flow via Hamilton-Jacobi equations for
generic quantum field theories leads always to a gradient flow for
$\beta$ functions \cite{deBoer:1999tgo}. Conversely a gradient flow for
$\beta$ functions is a pre-requisit for a holographic formulation of the RG flow.
The present approach may thus find applications in this direction, but
also in contexts beyond quantum field theory.

In order to extend the range of possible applications of the present
formalism it will be useful to generalise it towards non-marginal
couplings such as mass terms. Then, within mass dependent subtraction
schemes, the $\beta$ functions may depend explicitely on the scale(s)
which cases require further studies.

Finally the present approach requires as many scales $\tau_i$ as couplings
$g_a$. If this assumption is relaxed the reversibility of the matrices
of partial derivatives and/or the construction of a metric $\eta^{ab}$
imply constraints on the $\beta$ functions which merit further
investigations.

\section*{Acknowledgements}

The author acknowledges hosiptality of the University of California Santa Cruz
where this work was started, and support from the European Union's Horizon 2020
research and innovation programmes H2020-MSCA-RISE No. 645722
(NonMinimalHiggs).


\end{document}